# NMR and dielectric studies of hydrated collagen and elastin: Evidence for a delocalized secondary relaxation

Sorin A. Lusceac,[1] Markus Rosenstihl,[1] Michael Vogel,[1*]

Catalin Gainaru,[2] Ariane Fillmer,[2] Roland Böhmer[2#]

[1]*Institut für Festkörperphysik, Technische Universität Darmstadt, 64289 Darmstadt, Germany*

[2]*Fakultät für Physik, Technische Universität Dortmund, 44221 Dortmund, Germany*

Using a combination of dielectric spectroscopy and solid-state deuteron NMR, the hydration water dynamics of connective tissue proteins is studied at sub-ambient temperatures. In this range, the water dynamics follows an Arrhenius law. A scaling analysis of dielectric losses, 'two-phase' NMR spectra, and spin-lattice relaxation times consistently yield evidence for a Gaussian distribution of energy barriers. With the dielectric data as input, random-walk simulations of a large-angle, quasi-isotropic water reorientation provide an approximate description of stimulated-echo data on hydrated elastin. This secondary process takes place in an essentially rigid energy landscape, but in contrast to typical β-relaxations it is quasi-isotropic and delocalized. The delocalization is inferred from previous NMR diffusometry experiments. To emphasize the distinction from conventional β-processes, for aqueous systems such a matrix-decoupled relaxation was termed a ν-process. It is emphasized that the phenomenology of this time-honored, 'new' process is shared by many non-aqueous binary glasses in which the constituent components exhibit a sufficient dynamical contrast.

# 1. Introduction

The biological functions of hydrated proteins result from various complex dynamical processes [1]. Despite substantial progress, there is still a controversial debate about the nature and interplay of protein and water dynamics in these systems [2,3,4,5,6,7]. It was postulated that the protein is slaved [2] or plasticized [3] by the hydration water. Also, it is debated to what extent the slowdown of protein and water dynamics upon cooling can be understood in terms of a glass transition. Regarding water, the additional question arises whether the temperature-dependent molecular dynamics in the hydration shells of proteins can yield insights into the glass transition of the bulk liquid, which cannot be supercooled below 235 K due to interference of crystallization. Some workers argue that near 225 K protein hydration water exhibits a fragile-to-strong transition [8,9] as a consequence of a postulated underlying liquid-liquid phase transition [10], similar to that proposed for the deeply supercooled bulk liquid [11]. Challenging this conclusion for hydrated proteins, other workers put forward that the structural relaxation of water vanishes upon cooling [5] or that secondary processes are observed at low temperatures [4,5,12,13,14,15].

The phenomenology of the temperature-dependent dynamics of protein-water systems resembles that of other glass-forming mixtures composed of large and small molecules [7]. In such systems, it is often found that the dynamics of the components decouple upon cooling. While the motion of the large molecules freezes in on the experimental time scale at the glass transition temperature $T_g$, the small molecules can retain substantial mobility at $T < T_g$. It was often found that the temperature dependence of the dynamics of the small particles changes from a Vogel-Fulcher-Tammann- to an Arrhenius-type of behavior when crossing $T_g$ upon cooling. Examples for related decoupling phenomena are the transport of ions in inorganic glasses [16,17], the motion of guest molecules in amorphous matrices [18], and dynamically asymmetric mixtures, e.g., polymer blends [19,20]. The glassy slowdown of mixtures of small and large particles was also studied in theoretical work. Based on mode-coupling theory, it was argued that upon cooling the small particles show a crossover from a diffusion typical of liquids to a diffusion in a random potential produced by the large particles [21]. More recently, a molecular dynamics simulations study of a binary soft-sphere mixture reported a small-particle localization transition in the frozen glassy matrix formed by the large particles [22]. As a precursor of this transition, the small particles show anomalous diffusion, resembling the phenomenology predicted in theoretical studies of confined fluids [23] and the Lorentz gas [24].

Here, we study whether the dynamics of protein-water mixtures and of binary glass-forming liquids can be understood on common grounds. For this purpose, we investigate the dynamics of water in essentially static elastin and collagen matrices, i.e., below the glass transition or denaturation temperatures. We combine results from dielectric spectroscopy (DS) and from $^2$H NMR to monitor the rotational motion of water in broad ranges of time and temperature. In particular, performing random-walk (RW) simulations, we investigate whether both methods yield consistent results. In addition, we discuss literature results of pulsed-field gradient (PFG) NMR studies which probe translational motions of the water molecules on a micrometer length scale.

## 2. Theoretical background
### 2.1 Dielectric spectroscopy (DS)

For dielectric investigations the quantity of interest is usually the dielectric loss $\varepsilon''(\omega)$, i.e., the imaginary part of the complex dielectric permittivity $\varepsilon^*(\omega)$. Within the linear response regime, and neglecting cross-correlation factors, $\varepsilon''(\omega)$ is related to the reorientational dipole-dipole autocorrelation function $\phi_1(t) = \langle \cos\theta(0)\cos\theta(t) \rangle$ via [25]:

$$\varepsilon''(\omega) = \Delta\varepsilon\, \omega\, \mathrm{Re}\{F[\phi_1(t)]\}\,. \tag{1}$$

Here, F[..] denotes the Fourier transform and $\Delta\varepsilon$ is the dielectric strength of the relaxation process.

*2.2 Deuteron NMR*

In $^2$H NMR, we probe the quadrupolar contribution to the resonance frequency for deuterons in protein hydration water. This quadrupolar frequency is given by [26]

$$\omega_Q = \pm \frac{\delta}{2}\left(3\cos^2\theta - 1\right) . \qquad (2)$$

Here, the ± signs refer to the two allowed transitions of the I = 1 nucleus, δ denotes the anisotropy parameter of the quadrupolar interaction tensor, and θ is the angle between the direction of the O-$^2$H bond and the static external field $\vec{B}_0$. We exploit this orientation dependence of the $^2$H resonance frequency to study the reorientational dynamics of water.

For disordered materials, the anisotropy of the quadrupolar interaction together with a powder average usually results in broad Pake-like $^2$H spectra if the rotational correlation time of water exceeds the inverse anisotropy parameter, τ >> 1/δ ~ 1 μs. When the temperature is increased and the molecular dynamics becomes faster, these broad spectra collapse if τ ≈ 1/δ. Eventually, motionally averaged spectra result in the limit of fast motion, τ << 1/δ. In the presence of a broad distribution of correlation times, G(ln τ), dynamics in the limits of slow (τ >> 1/δ) and fast (τ << 1/δ) motion coexist in a certain temperature range. Then, so-called "two-phase" spectra − weighted superpositions of a broad and a narrow component − are usually observed [27].

Rotational motions with correlation times in the range of milliseconds can be studied via stimulated-echo decays [26,28]. In these experiments, a suitable three-pulse sequence (pulse-$t_1$-pulse-$t_m$-pulse-$t_2$) is used to generate a stimulated echo at a time $t_2 = t_1 << t_m$. The delays involved in the pulse sequence are called evolution time $t_1$, mixing time $t_m$, and detection time $t_2$. Stimulated-echo decays can be observed by varying the length of $t_m$ for a constant value of $t_p \equiv t_2 = t_1$. When appropriate pulse phases and lengths are applied, the rotational correlation function

$$F_2(t_p, t_m) = \left\langle \cos\left[\Phi(0, t_p)\right] \cos\left[\Phi(t_m + t_p, t_m + 2t_p)\right] \right\rangle , \qquad (3)$$

can be measured. Here, ⟨...⟩ denotes the ensemble average and $\Phi(t_a, t_b) = \int_{t_a}^{t_b} \omega_Q(t') \, dt'$ is the accumulated phase. When molecular dynamics during short evolution and detection times can be neglected, i.e., if $t_p \ll \tau$ and $t_p \ll t_m$, Eq. (3) reduces to

$$F_2(t_p, t_m) = \langle \cos[\omega_Q(0) t_p] \cos[\omega_Q(t_m) t_p] \rangle. \tag{4}$$

Molecular reorientations taking place during the mixing time $t_m$ will attenuate the stimulated-echo amplitude. In addition, spin-lattice relaxation (SLR) leads to a decay of the echo amplitude so that the corresponding relaxation time $T_1$ limits the experimental time window at long $t_m$. The length of the evolution time determines the angular resolution of the experiment, which is poor for small $t_p$ and high for large $t_p$. Therefore, information about the geometry of a reorientational process can be obtained if $F_2(t_p, t_m)$ is available for various evolution times $t_p$ [26,28]. More details about the interpretation of $^2$H NMR data for hydrated proteins can be found in the Appendix.

### *2.3 Random-walk (RW) simulations*

The calculation of $^2$H NMR observables using RW simulations is described in Refs. [28,29,30]. Briefly, a random-number generator is employed to generate a sufficiently large number of trajectories that describe the orientations of the relevant bonds as a function of time. Based on these trajectories, it is possible to calculate the time evolution of the corresponding frequencies $\omega_Q$, which in turn permits the computation of stimulated-echo data for comparison with experimental results. The input of RW simulations are, on the one hand, the geometry of the rotational motion, i.e., jump angles or preferred axes if an anisotropic process is to be simulated, and, on the other hand, the time scale of the dynamics as characterized by the correlation time $\tau$ or a potential distribution $G(\ln\tau)$.

For the present simulations, we start from a logarithmic Gaussian distribution

$$G(\ln \tau) = \frac{1}{\sigma\sqrt{2\pi}} \exp\left[-\frac{(\ln \tau - \ln \tau_m)^2}{2\sigma^2}\right], \qquad (5)$$

which is centered around the maximum position $\tau_m$ and exhibits a variance $\sigma^2$. Expressing this distribution in terms of a distribution of energy barriers g(E) with a standard deviation $\sigma_E$, one has $\sigma_E = T\sigma$ if $\tau_m$ follows an Arrhenius law. To reduce calculation time, we use a $3\sigma$ cutoff for G(ln$\tau$). With G(ln$\tau$) taken from DS, the only input for the RW simulations is the geometry of water's rotational motion. Thus, information about the mechanism governing the water dynamics can be gained when comparing the results from various motional models with the experimental data.

NMR stimulated-echo data are calculated following the procedure described in Ref. 30. We use $t_p$ times that are 2 µs longer than the experimental ones to account for the spin evolution during the finite pulse lengths. The protein contribution and the damping due to SLR are corrected for as described in the Appendix.

## 3. Experimental details

Lyophilized collagen (C) from bovine achilles tendon and elastin (E) from bovine neck ligament were obtained from Sigma-Aldrich. The samples were prepared by careful mixing of weighed amounts of protein and water. The compositions of the samples are specified by the hydration level h = $m_{H2O}/m_{C,E}$. The samples are designated as 'E' or 'C' followed by the effective h-value in percent. For the preparation of the NMR samples, we used E or C, which were dried over $P_2O_5$ in a dessicator for 3 days, and added $H_2O$ ($^1$H NMR, bi-distilled) or $D_2O$ ($^2$H NMR, Sigma-Aldrich). To avoid changes of the water content, the samples were sealed in the NMR tube using two-component epoxy adhesive (UHU, Bolton Group). The samples for the dielectric measurements were prepared similarly [14].

The experimental setups for the $^2$H NMR measurements were described elsewhere [31,32]. $^2$H NMR spectra were recorded using a solid-echo sequence and setting the length of the 90° pulses to 2 µs. Dielectric spectra were recorded employing a broadband α-spectrometer from Novocontrol in a frequency range $10^{-2}$ Hz < ν < $10^7$ Hz, together with an Agilent E4991A analyzer extending the range up to 1 GHz. For both setups parallel plate capacitors were used. The dielectric cells had geometrical capacitances of about 20 pF and were inserted into a Novocontrol cryostat in which the temperature was stabilized to within 0.2 K by a Quatro temperature controller.

## 4. Results

### *4.1 Dielectric spectroscopy*

Fig. 1 presents dielectric loss spectra, $\varepsilon''(\omega)$, of a C30 sample in a range exceeding 10 decades in frequency. In the spectra a pronounced relaxation peak is present. Over the entire range its shift with temperature, or that of the associated relaxation time $\tau = 1/(2\pi\nu)$, follows an Arrhenius law

$$\nu = \nu_0 \exp(-E_a/k_B T). \tag{6}$$

The corresponding time scales, as shown in Fig. 2, can be described by an activation energy $E_a = 0.55$ eV or 6330 K and a pre-exponential factor $\nu_0 = (2\pi\tau_0)^{-1} = 1\times10^{19}$ s$^{-1}$. The energy barrier is that required to break two hydrogen bonds. The large prefactor may indicate the existence of an entropy factor, i.e., if $E_a$ is considered a free energy $E_a = U - TS$ (with U the internal energy and S the entropy), then Eq. (6) is $\nu = \nu_{0,\text{eff}} \exp(-U/k_B T)$ and $\nu_{0,\text{eff}} = \nu_0 \exp(S/k_B)$. Alternatively, the modification of Eq. (6) may be rationalized by stating that the energy barrier is linearly temperature dependent.

A strong increase of $\varepsilon''$ on the low-frequency side of the relaxation peak, see Fig. 1, signals conductivity contributions. The time scales resulting from the loss peaks and the conductivity components do not follow the same temperature dependence in contrast to what is typically observed in the α-relaxation regime of *neat* supercooled liquids. For C30 the two spectral features overlap more at higher temperatures, as is also observed in the dielectric response of other hydrated systems [14,33]. There are also other aspects that distinguish the (hydration) loss peak of hydrated collagen from an α-process: The peak height (see Fig. 1) and also the dispersion step Δε [14] increase with temperature, opposite to what is typically observed for an α-process. Furthermore, unlike for a typical α-process, the loss peak is symmetrically broadened.

With the applicability of an Arrhenius law, the dielectric loss can be rescaled [34,35,36] to yield a distribution of energy barriers

$$g(E) = g(T \ln \nu_0/\nu) \approx \frac{2\varepsilon''(\nu)}{\pi T \Delta \varepsilon} \ . \tag{7}$$

The scaling works for sufficiently broad and temperature independent distributions g(E). The scaled dielectric data of hydrated collagen presented in Fig. 3(a) show that a master plot is obtained, justifying these assumptions. A similar conclusion was previously reached for hydrated elastin [14] and for later comparison with NMR data we reproduce its scaling plot as Fig. 3(b). The only adjustable parameter entering the scaling procedure is the prefactor $\nu_0$, cf. Eq. (7). We obtain $\nu_0 = 3\times10^{18}$ s$^{-1}$ for E46 and $\nu_0 = 1\times10^{19}$ s$^{-1}$ for C30. The mean values and the standard deviations of the resulting distributions g(E) can be determined by adapting Eq. (5) to the scaled data. As shown in Fig. 3, $E_a$ and $\sigma_E$ are quite similar for hydrated elastin and collagen.

Fig. 2 shows that for sufficiently large hydration levels, the water on the protein surface exhibits a quasi-universal dielectric relaxation. However, for h > 0.4 the loss peaks due to the hydration water at sub-zero temperatures is partially masked by a peak stemming from crystallized water [14]. It was demonstrated that this excess ice peak can selectively be suppressed by an electrical cleaning technique [37]. This made it possible to identify a 'critical' hydration level, corresponding to a complete "filling" of the protein's hydration shell, so that only non-crystallizable water covers the shell of the protein [14].

One may ask whether drying of an over-hydrated protein could also lead to a particular numerical value of the hydration level. Our drying experiments are documented in Fig. 4 for a collagen sample with an initial h = 0.7. Drying took place at room temperature within an unsealed dielectric cell. To check for the water content of the specimen it was occasionally and quickly cooled to 180 K at which a dielectric spectrum was recorded and immediately thereafter the sample was heated back to 300 K. The data in Fig. 4 show that within a few days the sample is dry in the sense that only structural water remains. The structural water is located inside this triple-helix protein [38] and gives rise to a faint dielectric loss. In Fig. 4 the dielectric spectra are labeled by approximate h levels which were estimated by comparison with data obtained for a wide range of hydrations [14]. Hence, at least in the present experiment, a particularly stable h value or range could not be identified. Obviously, at room temperature it is very easy for the hydration water to detach completely from the protein surface.

*4.2 Deuteron NMR*

Fig. 5 shows partially relaxed solid-echo spectra of E43, reflecting the water contribution to the NMR line shape, for its isolation from the protein water fraction see the Appendix. In a temperature range between about 170 and 220 K, we observe two-phase spectra comprised of a narrow Lorentzian and a broad Pake pattern. The relative intensity of both line-shape components continuously changes when the temperature is varied. Such a behavior indicates the existence of a broad distribution of correlation times $G(\ln\tau)$: As outlined in Section 2.2, the water molecules exhibiting fast ($\tau \ll 1$ μs) and slow ($\tau \gg 1$ μs) rotational motion contribute to a narrow line and a broad line, respectively [27]. Upon cooling, $G(\ln\tau)$ shifts to longer times and, consequently, the fraction of fast molecules becomes smaller and the intensity of the narrow line decreases continuously. Thus, the NMR line-shape behavior confirms the existence of dynamical heterogeneities, which was tacitly assumed when determining $g(E)$ from the dielectric loss, see Eq. (7). The fits included in Fig. 5 demonstrate that the narrow line-shape component exhibits a Lorentzian shape. Such a Lorentzian line is found not only for the hydration water of elastin, but also for that of collagen, as can be inferred from the spectrum for C25. Observation of a Lorentzian line implies the existence of an isotropic water reorientation, leading to complete averaging of the anisotropy of the NMR interactions. More precisely, in the studied temperature range, any anisotropy of the water dynamics is too weak to retain a residual line width that would be recognizable from the motionally averaged spectrum.

Fig. 6 shows the normalized fraction $W_{fast}$ of the Lorentzian line to the total spectral intensity for the studied temperatures. We see that $W_{fast}$ smoothly increases when the temperature is increased, indicating the existence of a continuous distribution $G(\ln\tau)$. The data from DS imply that, using Eq. (6), the distribution $G(\ln\tau)$ can be derived from a temperature independent distribution of activation energies $g(E)$. In this case, $g(E)$ can be obtained from $W_{fast}(T)$ according to [27]

$$\frac{dW_{fast}(T)}{dT} = g(E)\ln\left(\frac{\tau^*}{\tau_0}\right). \qquad (8)$$

Like above, $\tau_0$ is the inverse of the attempt frequency and $\tau^*$ is the correlation time at which motional narrowing is observed ($\tau^* \approx 1/\delta$). Assuming a Gaussian distribution $g(E)$ and using $\tau_0 = 1/(2\pi\times10^{19})$ s as obtained from DS, together with $\tau^* = 10^{-6}$ s, it is possible to determine the parameters of the distribution by fitting $W_{fast}(T)$. In Fig. 6, we see that this approach, with the parameters $E_a = 6270$ K and $\sigma_E = 440$ K, enables a good interpolation of the experimental data.

For later analysis, we reproduce previously published stimulated-echo decays $F_2(t_p,t_m)$ of E43 at various temperatures in Fig. 7 [12]. They were measured in fully relaxed experiments and, thus, comprise contributions of water and of protein deuterons, see the Appendix. One recognizes that the initial decays due to water reorientation exhibit a pronounced stretching, consistent with a continuous distribution $G(\ln\tau)$, as obtained in the analysis of the solid-echo spectra.

### *4.3 Random-walk simulations of stimulated echoes*

Considering the previous finding of a crossover from isotropic to anisotropic reorientation for myoglobin hydration water upon cooling to below about 210 K [31], it is worthwhile to check whether similar effects arise for hydrated elastin. Therefore, we study the motional mechanism in E43 at low temperatures. For this purpose, we compare stimulated-echo results at 145 K ≤ T ≤ 165 K [12] with simulations for several models of water's molecular motion. With the distribution of correlation times and its temperature dependence taken from DS, only the reorientational geometry remains to be specified for the present RW simulations.

To describe our stimulated-echo data, we test three models: a two-site jump (2SJ), a four-site jump (4SJ), and an isotropic random jump (IRJ). Use of the latter model is motivated by the observation of a Lorentzian component in the two-phase spectra. The considered 2SJ is a $\pi$ flip about the $C_2$ symmetry axis of the water molecule, which is found in several crystalline hydrates [39]. Moreover, a $\pi$ flip was observed in a recent MD simulation study on elastin hydration water at sufficiently low temperatures [40]. For the 2SJ, we use an angle of 104.5° between the possible orientations of an O-H bond, which is found for the H-O-H bond angle of the water molecule in the gas phase. The used 4SJ is a tetrahedral jump, as reported for water molecules in hexagonal ice [41] and in zeolites [42]. Hence, the 2SJ and the 4SJ models involve large jump angles of the order of the tetrahedral angle. In previous work [12], the existence of large-angle jumps was inferred by analyzing the evolution-time dependence of $F_2(t_p,t_m)$.

In Fig. 7(a), we compare the stimulated-echo decay measured at 165 K for an evolution time of $t_p = 30$ μs with the simulated curves for the 2SJ and 4SJ jump models. The simulated decays strongly depend on the employed model. We see that the decay is steeper when more sites are available for an O-$^2$H bond. Such behavior is expected since $F_2(t_p \to \infty, t_m \to \infty) = 1/N$ if jumps occur between N equally populated, magnetically inequivalent sites. In the experiment, a plateau value could not be observed since SLR leads to an additional damping of the stimulated-echo amplitude for $t_m > 10^{-1}$ s. In the simulation, this effect is mimicked by multiplication of the calculated data with the experimental SLR decay, see Appendix. Moreover, it is evident from Fig. 7(a) that the decays for the 2SJ and the 4SJ jump models do not match the experimental data. The large difference means that more sites are involved in water reorientation. In particular, the IRJ model enables a reasonable reproduction of the experimental curve.

However, we obtain comparable decays when assuming slightly distorted 2SJs and 4SJs as follows: We redefined our models by allowing orientations on 2 and on 4 cones, respectively, where the axes of the cones are the allowed directions of the undistorted 2SJ and 4SJ models, see Fig. 7. Each jump is a large-angle jump between different cones and the new position is randomly chosen around the exact orientation of the 2SJ or 4SJ models. Stimulated-echo decays resulting from models for a cone semi-angle $\chi = 10°$ are included in Fig. 7(a). We see that the decays are comparable to that obtained within the IRJ model. Hence, the experimental stimulated-echo decays do not allow us to discriminate between isotropic jumps and distorted N-site jumps, provided the latter exhibit sufficient distortion of the molecular orientations, i.e., $\chi \geq 10°$.

Next, we investigate whether the temperature dependence of $F_2(t_p,t_m)$ is consistent with expectations on the basis of the DS data. For this purpose, we use the distribution $G(\ln\tau)$ determined in DS for the studied temperatures and simulate the resulting stimulated-echo decays for hydrated elastin. Here, we focus on the model of a 4SJ with a distortion $\chi = 10°$, but comparable results are obtained for other isotropic or distorted large-angle jump models. In Fig. 7(b), we see that the simulated and the experimental data show comparable temperature dependence. Hence, DS and NMR yield consistent results, when it is assumed that the O-H bonds perform large-angle jumps involving a sufficiently large number of orientations, $N \geq 4$ (which also includes the distorted 2SJ). At T < 165 K, the short-time decay of $F_2(t_p,t_m)$ is slightly more pronounced than expected on the basis of the dielectric distribution $G(\ln\tau)$, see Fig. 7(b). Two possibilities come to mind to rationalize these deviations: (i) In part, they could be due to uncertainties regarding the exact magnitude of the protein contribution at the studied temperatures [12]. (ii) This discrepancy could also be taken to suggest that there is a somewhat faster dynamical process which is not observable in DS, e.g., a very weakly distorted $\pi$ flip about the $C_2$ symmetry axis of the water molecule. For the hydration water of myoglobin [51], NMR data implied the existence of such type of low-temperature water motion. For hydrated elastin and collagen this possibility is unlikely since Fig. 3 shows that there is an excellent agreement of the energy barrier distribution from the $T_1$ and $\varepsilon''$ data over the entire temperature range, see also Section 5.1, below [43]. It is clear, however, that if the cone angle $\chi$, which describes the deviations from an ideal 2SJ, decreases continuously with decreasing temperature [44], this will lead to a decrease of $\Delta\varepsilon$ as experimentally observed [14].

## 5. Discussion

### 5.1. Comparison of DS and NMR time scales for hydrated elastin

In the present and in previous studies [12,14,31], DS and NMR methods were used to investigate the dynamics of elastin hydration waters. Therefore, the question arises whether these methods yield consistent results. Analysis of NMR SLR requires knowledge about the shape of the spectral density $J(\omega)$ and, thus, about the shape of $G(\ln\tau)$. Due to lack of information, previous $^2$H NMR work compared results for symmetric and asymmetric distributions $G(\ln\tau)$ [12]. When using the asymmetric Cole-Davidson spectral density for SLR analysis, correlation times $\tau_a$ were obtained, which significantly deviated from an Arrhenius law, while an analysis on the basis of the symmetric Cole-Cole spectral density provided correlation times $\tau_s$, which are consistent with those from DS, see Fig. 2.

Given an Arrhenius law is obeyed, the underlying distribution $g(E)$ can be obtained from the temperature dependence of $T_1$ without recourse to a specific model. Identifying $\nu$ in Eq. (7) with the Larmor frequency and recognizing that by virtue of the fluctuation-dissipation-theorem $\varepsilon''/T$ is proportional to $1/T_1$ [45], we obtain

$$g(T \ln \nu_0/\nu_{Larmor}) \approx \frac{K}{T_1 T} \qquad (9)$$

where $K$ is a proportionality constant. Using this relation, we can include the $^2$H SLR times $T_1$ of E43 and C25 [12] in the scaling plot of the DS data, Fig. 3(a) and (b), respectively. We see that the scaled $T_1$ values are compatible with the scaled dielectric data. Furthermore, the distribution of activation energies g(E) was determined from the $^2$H NMR line shapes of E43 using Eq. (8). The resulting distributions are also compatible to that obtained from scaling of the dielectric data, see Fig. 3(b).

In previous work [12], the stimulated-echo decays of E43 were fitted to a Kohlrausch function, $\exp[-(t/\tau)^\beta]$, in order to determine mean correlation times. As Fig. 2 shows, the resulting average NMR time constants are about 1-2 orders of magnitude longer than those obtained from DS and they show a somewhat smaller activation energy, $E_a = 0.45$ eV. Several effects contribute to this difference: (i) The mean correlation times determined in NMR and DS reflect averages on a linear and on a logarithmic scale, respectively, which differ by orders of magnitude in the present case of a very broad distribution of correlation times [46,47,48]. (ii) The NMR data were analyzed using a Kohlrausch function, which does not correctly describe the shape of decays resulting from a logarithmic Gaussian distribution $G(\ln\tau)$. (iii) The correlation function of water dynamics is highly stretched and its decay range exceeds the time window of stimulated-echo studies. Thus, the observed discrepancies cannot be taken as evidence that NMR and DS probe different dynamical processes. This assertion is confirmed by the reasonable agreement of measured and simulated stimulated-echo decays in Fig. 7(b), as was discussed near the end of Section 4.2.

### 5.2. Comparison of DS and NMR data for dynamics of hydration waters in various protein matrices

In recent years, DS and NMR were applied to investigate the temperature-dependent dynamics for a number of hydrated proteins. Here, we studied water dynamics in essentially rigid protein matrices, i.e., *below* the glass transition temperature of elastin ($T_g = 303$ K [49]) or *below* the denaturation temperature of collagen ($T_d = 370 - 390$ K [50]). In rigid matrices, water dynamics was found to be describable by an activation energy $E_a \approx 0.5$ eV [4,5,14,15,19], consistent with the present results. This "universal" activation energy of water dynamics was not only observed for proteins, but for various kinds of matrices, including silica matrices [4,5,19]. For water dynamics *above* the glass transition of a mixture, deviations from an Arrhenius behavior and additional relaxation processes were reported [5,15,19], but this temperature range is not covered in the present study.

For the hydration waters of elastin, collagen, and myoglobin [31,51], a Lorentzian line is found as $^2$H NMR spectrum down to T ≈ 230 K. Thus, water dynamics is isotropic in this temperature range, independent of the value of $T_g$. At least, any potential anisotropy is too weak to be observed in the NMR line shape. Upon further cooling, the motionally narrowed component in the spectra of elastin's hydration water yields no evidence for deviations from a Lorentzian shape and, hence, from an isotropic motion. For hydrated myoglobin, on the other hand, NMR line-shape analysis indicated a crossover from isotropic to anisotropic reorientation in the temperature range between about 230 and 200 K [31,51]. Specifically, for hydrated myoglobin, it was observed that the narrow component of the two-phase spectrum changes its shape from Lorentzian to non-Lorentzian. We cannot completely exclude that there are also some changes in the cone angle $\chi$ characterizing the reorientation of elastin hydration water at lower temperatures, see Sec. 4.3. Such a difference of line-shape behaviors between hydrated elastin and hydrated myoglobin would then suggest that the mechanism for the reorientation is somewhat different for the hydration waters of the various proteins below about 200 K.

Previously, $^1$H PFG NMR was used to investigate translational motions of various protein hydration waters [52,53], including hydrated collagen [54] and hydrated elastin [55]. For these proteins, diffusivities of D = (1 ... 2)×10$^{-10}$ m$^2$s$^{-1}$ were found to characterize the motion of hydration water at ambient temperature. These values are about an order of magnitude smaller than the corresponding diffusivity of bulk water and provide clear evidence for the existence of a translational motion of protein hydration waters on a micrometer length scale. In particular, for hydrated elastin and collagen, the results of the PFG NMR studies [54,55] demonstrate that long-range water transport occurs in basically rigid protein environments and, hence, that the water dynamics is non-local.

*5.3 Decoupled water motion*

With the terminology known from glass forming liquids, the motion of the hydration water was variously designated as a primary or as a secondary relaxation process. Focusing on the present case of hydration waters of connective tissue proteins, an assignment in term of an α-relaxation can be ruled out. The process that we observe exhibits neither non-Arrhenius characteristics nor is its relaxation spectrum asymmetrically broadened. Furthermore, the temperature dependence of its dielectric relaxation strength is different from that typically observed for an α-process of supercooled liquids. Hence, it is suggestive to identify the observed, symmetrically broadened relaxation with a β-process. However, there are several observations demonstrating that the hydration water dynamics in collagen and elastin is not of the Johari-Goldstein (JG) β type [56]. Firstly, at temperatures much below the glass transition or denaturation temperatures, the NMR spectrum is given by a narrow Lorentzian. This feature would be hard to reconcile with the fact that typical JG β-relaxations are associated with anisotropic motions. Secondly, hydration water can perform a translational displacement on the micrometer length scale deep in the glassy phase [54,55], whereas conventional β-processes are understood to involve only spatially localized molecular motions.

However, the scenario referred to so far does not capture the situation in various *binary* glass formers, particularly in mixtures composed of constituents exhibiting a high dynamical contrast [19,20,57,58]. Here, the slower component can form a more or less rigid matrix in which the faster, and usually molecularly smaller component performs translational jumps. Well-known substances involving a decoupled motion in a rigid energy landscape are plasticized oligo- and polymer systems as well as fast ion conductors. A good example for the former is benzene in oligo-styrene [59,60]. For these binary mixtures, NMR measurements carried out well below $T_g$ revealed that the reorientation of the sixfold axes of the benzene molecules results in two-phase spectra and involves isotropic large-angle rotational jumps. The latter observation was ascribed to the presence of a translational diffusion process. In a disordered environment, the orientations of a mobile probe molecule are not expected to be strongly correlated as the molecule hops from site to site [61].

The present case of water motion on the surface of a strongly dynamically hindered protein provides similar conditions. With the observation of a translational diffusion process and large-angle rotational jumps, indeed a similar phenomenology is found. Furthermore, the observed activation energy equals that required to break two hydrogen bonds being a precondition for a translational motion of water. Eventually, the water molecules may detach completely from the protein surface, thus rationalizing the drying evident in Fig. 4. Hence, the phenomenology of the relaxation of protein hydration water, which is documented in the present article, resembles that of the faster component of binary glass-forming mixtures, which are composed of constituents showing largely different molecular mobilities. While the phenomenology of the process is similar to that of a wide class of binary systems, it differs from that of the α- and β-process in neat systems, justifying to call it a 'new' process or ν-process, following previous work [62]. One of the main characteristics of the ν-process is that it is associated with long-range transport of fast particles, which is decoupled from the α-process of slow particles, although the latter can be strongly affected, e.g., in terms of plasticizing. However, the ν-process may be coupled to a local β-process of the slower particles, which may, e.g., affect the lifetime of pathways or inter-component hydrogen bonds. For different binary glass-forming mixtures, quantitative differences of the ν-process can result because depending on the composition of the mixture, the fast molecules may be isolated or located in clusters. In that sense, one may conceive a continuous variation of behavior between the limiting cases of a β- and a ν-type of process, implying that a strict distinction of the two may be difficult in some cases.

## 6. Summary and conclusions

We studied the dynamics of hydrated elastin and collagen for temperatures between 270 and 145 K, i.e., below the glass transition and denaturation temperatures of these proteins. In this entire range the motional correlation times of water strictly follow an Arrhenius temperature dependence with an activation energy of (6210 ± 120) K. This value corresponds to the energy required to break two hydrogen bonds. Via a scaling analysis of dielectric losses and of spin-lattice relaxation times, the distribution of activation energies could be described via a Gaussian profile with a standard deviation of (480 ± 20) K. A similar width was found for hydrated elastin by evaluating so-called two-phase deuteron NMR spectra. These are composed of a sharp Lorentzian and a broad Pake pattern, which are equally intense near 200 K indicative for a typical reorientational correlation time of $\delta^{-1} \approx 1$ μs.

Based on these temperature-dependent time scales and their distribution, we performed random-walk simulations using various models for the motional geometry of the water molecules. Among them are approaches involving two-site and four-site jumps, including distorted tetrahedral models, as well as an isotropic random jump. Substantially distorted N- site jumps and isotropic random jumps gave good agreement with the results from the deuteron stimulated-echo experiments. This shows that by an appropriate analysis of the NMR data good compatibility with the dielectric data can be obtained. No discontinuous transitions of any kind are required to explain the experimentally observed dielectric and NMR results discussed in the present article.

The symmetric distribution of correlation times and their Arrhenius temperature dependence indicate that the water motion takes place in an essentially rigid energy landscape, precluding an interpretation in terms of a structural (or α-) relaxation. Nevertheless, the presently observed dynamics should also not be confused with that of a typical JG β-relaxation because the motional process involves both a quasi-isotropic reorientation and a translational diffusion of the water molecules. The latter, distinguishing feature was directly demonstrated via NMR diffusometry for the rigid-matrix phases of hydrated elastin and hydrated collagen [54,55]. We point out that the phenomenology for this kind of secondary, but delocalized ν-process is by no means restricted to binary *aqueous* glasses. Examples from the literature, provided to underscore this statement, include plasticized oligo- and polymers, polymer blends, and fast ion conductors. For future studies it remains to explore if and to what extent the ν-process, as exemplified by the present results, is compatible with the dynamics as studied by the simulations of a binary soft-sphere system mentioned in the Introduction. On the experimental side, the challenge will be to identify classes of systems that allow for a more or less gradual variation from the limit of a localized (β-) to that of a delocalized (ν-) process, e.g., by variation of pressure or composition variables.

**Appendix**

In previous $^2$H NMR work on E43 and C25 [12], we found that water deuterons and protein deuterons coexist because of chemical deuteron-proton exchange. While both deuteron species exhibit common SLR behavior at low temperatures, the SLR of the protein deuterons is substantially longer than that of the water deuterons at higher temperatures, T ≥ 175 K, then enabling suppression of the protein contribution in partially relaxed experiments. In these experiments, we saturate the magnetization and start the experiment after an appropriately chosen delay that allows for recovery of the water contribution, but not for recovery of the protein contribution. Here, we exploit the possibility of successfully suppressing the protein signal in partially relaxed experiments on E43 at T ≥ 175 K to single out the line-shape contribution of hydration water. However, in stimulated-echo experiments at lower temperatures, both deuteron species contribute to the experimental data because of comparable SLR behaviors. At 165 K ≥ T > 155 K, two contributions to the magnetization recovery are still discernible in SLR experiments. Therefore, the stimulated-echo decay can be written as:

$$F_2(t_p, t_m) = p_W \, F_W(t_p, t_m) \exp\left[-\left(\frac{t_m}{T_{1W}}\right)^{\beta_W}\right] + p_P \, F_P(t_p) \exp\left[-\left(\frac{t_m}{T_{1P}}\right)^{\beta_P}\right]. \qquad (10)$$

Here, $p_W$ and $p_P = (1 - p_W)$ are the fractions of deuterons belonging to water and protein molecules, respectively. $F_W$ and $F_P$ are their respective contributions to the stimulated echo. Motivated by findings in $^{13}$C NMR [63], we assume that $F_p$ does not depend on the mixing time $t_m$. For both components, we describe the SLR using a Kohlrausch function with the parameters $T_{1W,P}$ and $\beta_{W,P}$. In our investigation, they are determined *independently* via SLR experiments. At temperatures T ≤ 155 K, we do not observe discernible SLR behaviors of water and protein deuterons so that $T_{1W} = T_{1P}$ and $\beta_W = \beta_P$ in Eq. (10).

**FIGURES AND CAPTIONS**

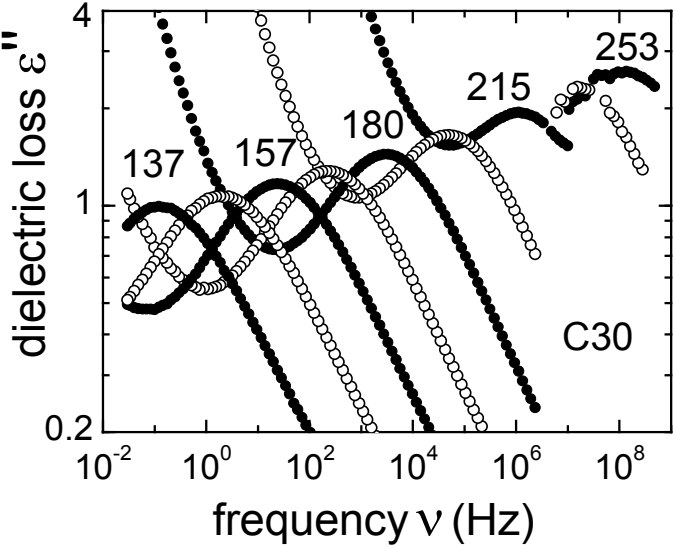

Fig. 1

Dielectric loss spectra of C30 recorded at various temperatures given in Kelvin.

28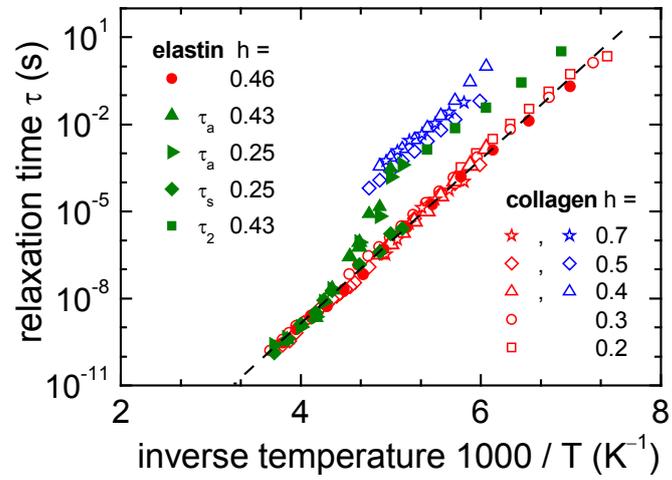

Fig. 2

Arrhenius plot of time constants for hydrated elastin (closed symbols) and hydrated collagen (open symbols). Green symbols are from NMR [12], the other symbols are from dielectric spectroscopy [14] and reflect the behavior of the hydration water (red symbols) and of excess ice (blue symbols). The dashed line is an Arrhenius law representing an activation energy of 6330 K.

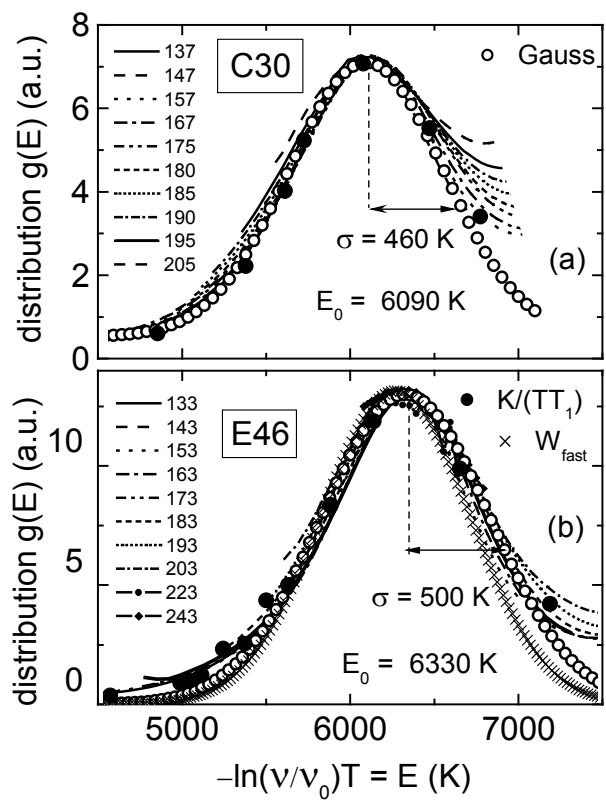

Fig. 3

The dielectric spectra of (a) C30 and (b) E46, represented as lines, comply with a scaling based on a thermally activated dynamics over a wide range of temperatures (given in Kelvin). Similar Gaussian distributions of activation barriers (represented by the open circles) are obtained for hydrated collagen and hydrated elastin. In each case, one parameter $\nu_0$ is sufficient to collapse all dielectric data. The solid circles represent $T_1$ measurements for C25 and E43 at a Larmor frequency of 76.8 MHz scaled using the values of $\nu_0$ obtained from the dielectric data. The crosses in frame (b) are from the temperature derivative of $W_{fast}$ for E43, which was determined via an analysis of two-phase NMR spectra.

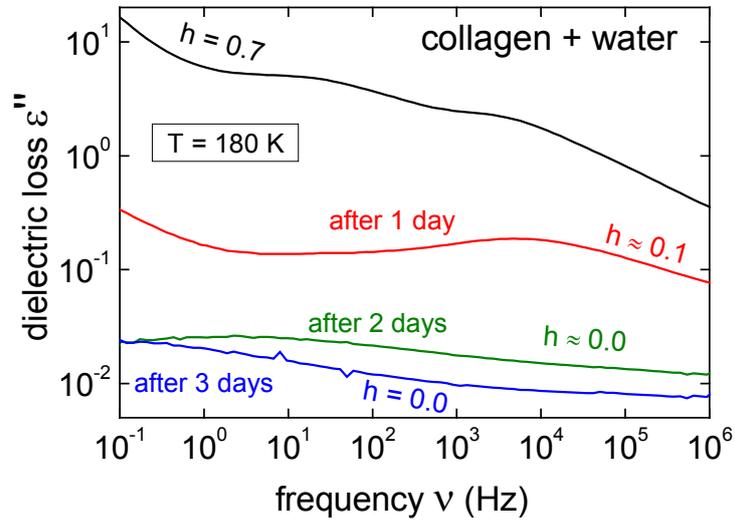

Fig. 4

Dielectric loss spectra of C70 measured at 180 K after the unsealed sample was kept at room temperature for the time intervals indicated in the figure.

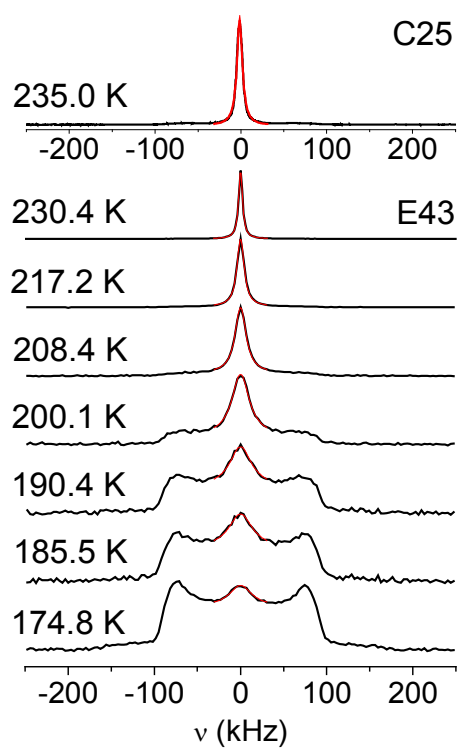

Fig. 5

Partially relaxed $^2$H NMR solid-echo spectra of E43 at the indicated temperatures. The delay between the saturation sequence and the solid-echo sequence was set to $T_{1W}$ = 3 - 40 ms, in harmony with the conditions outlined in the Appendix. For comparison, the fully relaxed $^2$H NMR solid-echo spectrum of C25 at T = 235 K is included. Close inspection of the latter reveals that in addition to the narrow water signal, there is a broad line resulting from protein deuterons, see Appendix. The red lines are fits of the narrow spectral components with a Lorentzian. All solid-echo spectra were recorded using a solid-echo delay of 20 μs.

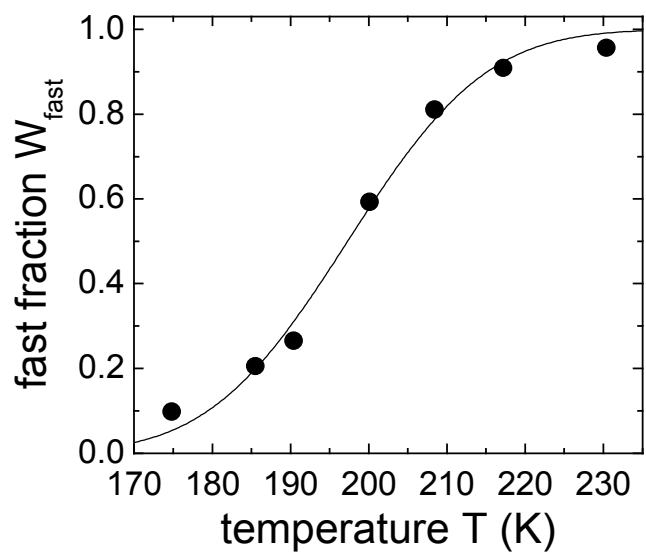

Fig. 6

Normalized fraction $W_{fast}$ of the Lorentzian line to the total spectral intensity for E43 as obtained from an analysis of the data shown in Fig. 5. The line is a fit with the integral of a Gaussian function.

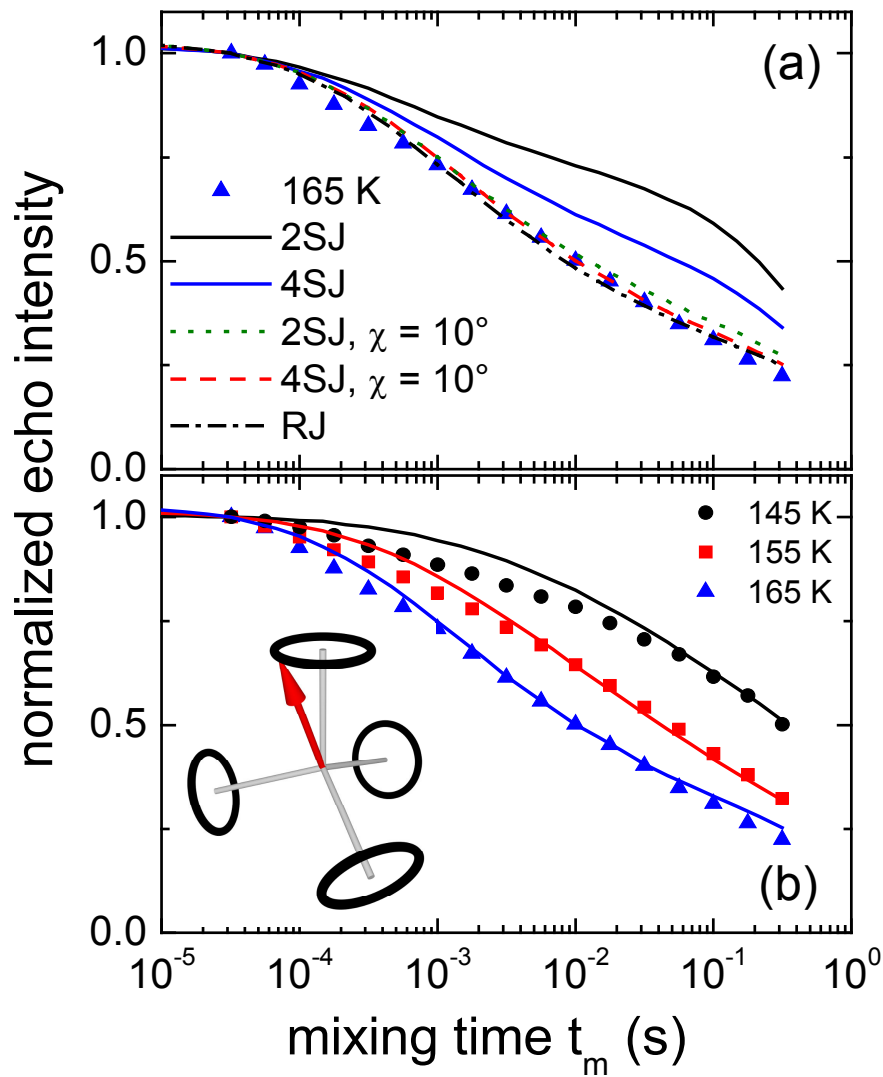

Fig. 7

$^2$H NMR stimulated echo decays for an evolution time of $t_p = 30$ μs. (a) Points are measured data at 165 K, lines are simulations using different models (see text for details). (b) Experimental stimulated-echo decays (symbols) for several temperatures together with the corresponding simulations (lines) for the distorted 4SJ model with $\chi = 10°$. All curves are normalized to the point at $t_m = 32$ μs. The inset in frame (b) is a sketch of the distorted 4SJ: The thick black lines symbolize the edges of cones which in turn represent the infinitely many orientations of an O-$^2$H bond permitted by the model.